# HYDRODYNAMIC MODEL OF FUKUSHIMA-DAIICHI NPP INDUSTRIAL SITE FLOODING

V.N.Vaschenko[1], V.I.Skalozubov[2], T.V.Gerasimenko[3], B. Vachev[4]

[1]*State ecology academy of Ukraine, V.Lypkivsky str, 35, Kyiv, 03035*
[2]*Institute for problems of NPP safety of NAS of Ukraine, Kyiv, daniilko@mail.ru*
[3]*International expert of UNO on climate change and energy saving*
[4]*Institute for Nuclear Research and Nuclear Energy, Bulgarian Academy of Sciences, Bulgaria, BG-1784, Sofia, Tzarigradsko chaussee, Blvd., 72*

While the Fukushima-Daiichi was designed and constructed the maximal tsunami height estimate was about 3 m based on analysis of statistical data including Chile earthquake in 1960. The NPP project industrial site height was 10 m. The further deterministic estimates TPCO-JSCE confirmed the impossibility of the industrial site flooding by a tsunami and therefore confirmed ecological safety of the NPP. However, as a result of beyond design earthquake of 11 March 2011 the tsunami height at the shore near the Fukushima-Daiichi NPP reached 15 m. This led to flooding and severe emergencies having catastrophic environmental consequences.

This paper proposes hydrodynamic model of tsunami emerging and traveling based on conservative assumptions. The possibility of a tsunami wave reaching 15 m height at the Fukushima-Daiichi NPP shore was confirmed for deduced hydrodynamic resistance coefficient of 1.8. According to the model developed a possibility of flooding is determined not only by the industrial site height, magnitude and distance to the earthquake epicenter, but also by duration of seismic impacts, power dissipation conditions, epicenter size and other conditions.

<u>Keywords</u>: flooding, tsunami, earthquake, environmental safety, hydrodynamic model

## The basic principles and assumptions of the calculation model for the Fukushima-Daiichi industrial site flooding

1. Earthquake with maximal acceleration $a_{eq}$ at depth $h_{eq}$ and at surface area $S_{eq}$ (fig.1) is the main source of perturbation wave in the water volume (in our case - the ocean) of height $h(t)$, which in general case may propagate in all directions $R(t)$ from the epicenter in horizontal plane ($t$ - is the current time of the perturbation wave propagation).

2. The perturbation wave propagation (the tsunami) is subject to seismic influence strength $F_{se}$, hydrodynamic resistance $F_{we}$ and wind pressure of density $\rho_{we}$ and speed $w_{we}$ ($F_{we}$) for in general case variable ocean depth $h_g(t)$, which is determined by specific bottom relief $f_p(R)$.

3. The perturbation wave propagation is assumed to be isothermic process and the sea water compression may be neglected, i.e. water density $\rho$ is constant.

4. The conservative assumptions are the following:
- the wind direction with $\rho_{we}$ density is moving in the same direction as the tsunami;
- as soon as the tsunami reaches the water breaker the water flow is almost instantly stops in horizontal direction and an deceleration wave height $h(R=L_0)$ abruptly increases;

- the deceleration wave height completely determines industrial site flooding condition but without considering hydrodynamic and dissipative processes in between the water breaker and the industrial site.

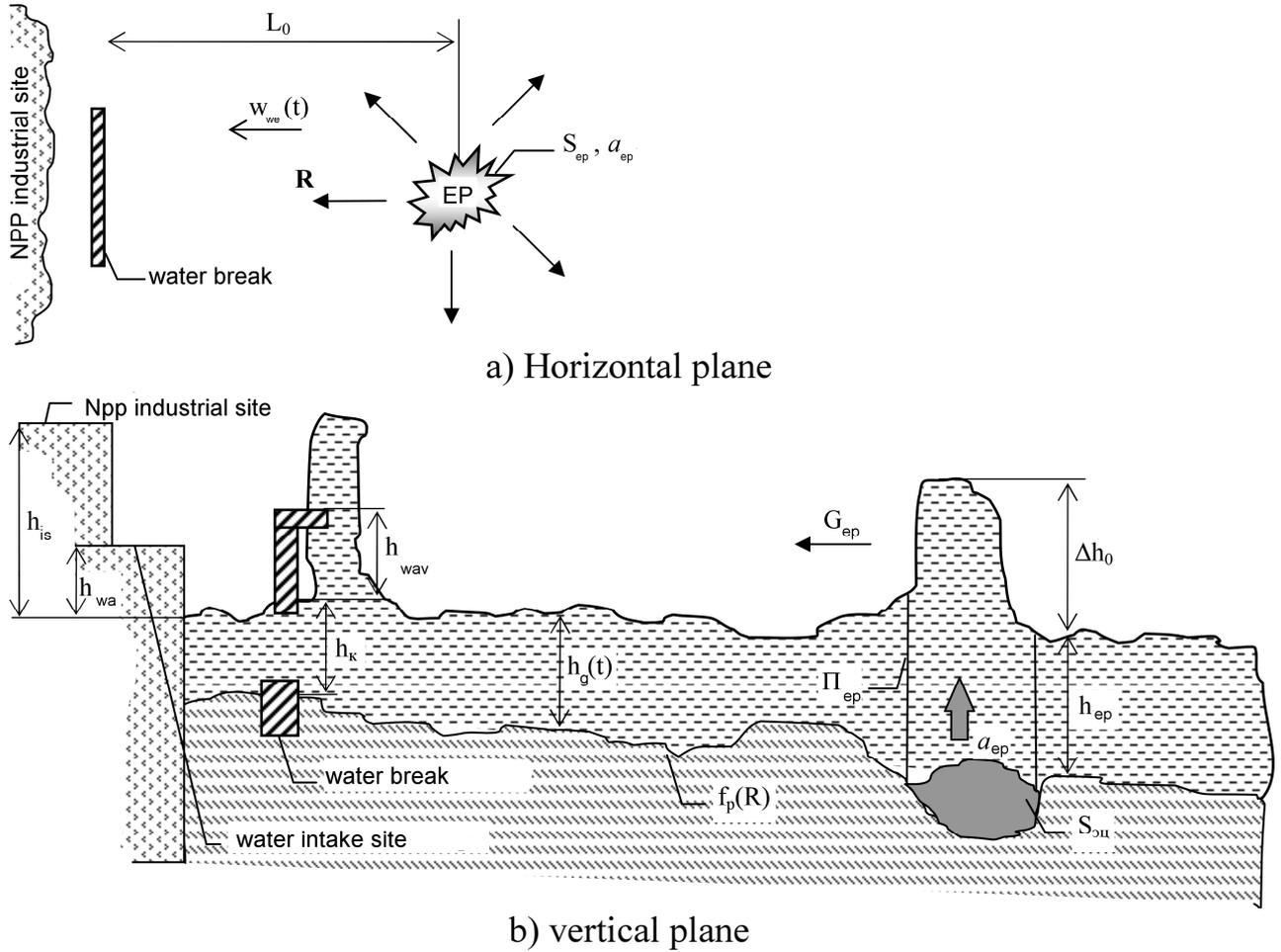

a) Horizontal plane

b) vertical plane

Fig.1. Calculation model for Fukushima-Daiichi NPP industrial site flooding.

Te perturbation wave development process caused by the earthquake is figuratively split into two temporal stages:

1st stage - excitation of the wave by the earthquake epicenter ($0 \leq t \leq \Delta t_c$);

2nd stage of the perturbation wave propagation in horizontal plane ($\Delta t_c \leq t \leq t_0(R=L_0)$);

where $\Delta t_c$ is the tsunami excitation time.

Such modeling is acceptable in case that the tsunami excitation time $\Delta t_c$ is much smaller than than its propagation to the shore $t_0$.

Conservative condition for the NPP industrial site flooding is

$$\Delta h / h_{pr} \geq 1 \qquad (1)$$

Then within the suggested framework the model of the wave excitation with height $h = h_{ep} + \Delta h$ at the earthquake epicenter (1st stage):

$$\rho S_{ep} \frac{dh}{dt} = -G_c, \qquad (2)$$

$$\rho S_{ep} \frac{d}{dt}\left(h\frac{dh}{dt}\right) = \rho S_{ep} a_{ep} h(t) - \rho v \Pi_{ep} \frac{dh}{dt} \qquad (3)$$

with boundary conditions

$$h(t = 0) = h_{ep}, \qquad (4)$$

$$\frac{dh}{dt}(t = 0) = 0, \qquad (5)$$

$$h(t = \Delta t_c) = h_{ep} + \Delta h_0, \qquad (6)$$

$$\frac{dh}{dt}(t = \Delta t_c) \to 0, \qquad (7)$$

where $G_c$ is the perturbation wave (the tsunami) spread from the epicenter; $v$ - the fluid kinematic viscosity coefficient; $\Delta t_c$ is the total duration of the underground shocks; $\Pi_{ep}$ - is the perimeter of the water "column" perturbation.

After transformation the equations of the tsunami excitation at the 1st stage (2) and (3) with boundary conditions (4)-(7) will have the form:

$$h\frac{d^2 h}{dt^2} = a_{ep} h - \frac{\Pi_{ep}}{S_{ep}} v \frac{dh}{dt} - \frac{G_c^2}{\rho^2 S_{ep}^2}, \qquad (8)$$

$$\frac{dh}{dt} = -\frac{G_c}{\rho S_{ep}}. \qquad (9)$$

The final result of equations (4)-(9) integration is the tsunami wave spread in the horizontal plane $G_c(t=\Delta t_c)$ and initial height of the propagation wave $\Delta h(t=\Delta t_c)$.

The model of the perturbation wave propagation with height $\Delta h(t)=h-h_g$ at the 2nd stage has the form $(t>\Delta t_c)$:

$$\rho \frac{d}{dt}\left[(\Delta h + h_g) R^2\right] = 0, \qquad (10)$$

$$\rho \frac{d}{dt}\left[(\Delta h + h_g) R^2 \frac{dR}{dt}\right] = F_{we}(t) - F_{hy}(t), \qquad (11)$$

with boundary conditions:

$$h(t = \Delta t_c) = \Delta h_0, \qquad R(t = \Delta t_c) = R_{ep}, \qquad (12)$$

$$\frac{dh}{dt}(t = \Delta t_c) = 0, \qquad (13)$$

$$\frac{dR}{dt}(t = \Delta t_c) = \frac{G_c(t = \Delta t_c)}{\rho S_{ep}}, \tag{14}$$

$$\frac{dR}{dt}(R = L_0) \to 0, \tag{15}$$

$$h_g(t) = f_p(R), \quad h_g(R = 0) = h_{ep}, \quad h_g(R = L_0) = h_к, \tag{16}$$

where deduced current forces influencing the wave of the wind of the same direction $F_{we}$ and hydrodynamic resistance $F_{hr}$

$$F_{we} = \rho_{we} w_{we}^2 \Delta h R, \tag{17}$$

$$F_{hy} = \lambda_{hy} \rho h R \left(\frac{dR}{dt}\right)^2, \tag{18}$$

then the calculation model in criterial form:

$$T_1 = \frac{t_1}{\Delta t_c}, \quad H = \frac{h}{h_{ep}}, \quad \Delta H = \frac{\Delta h}{h_{ep}}, \quad H_г = \frac{h_g}{h_{ep}},$$

$$\overline{\rho} = \frac{\rho_{we}}{\rho}, \quad L = \frac{R}{L_0}, \quad T = \frac{t}{M_t}, \quad \overline{w} = \frac{w_{we}}{M_w}, \quad \overline{G} = \frac{G_c}{M_G}.$$

at the 1st stage

$$\frac{dH}{dT_1} = -K_G \overline{G}, \quad K_G = \frac{M_G \Delta t_c}{\rho S_{ep} h_{ep}},$$

$$H \frac{d^2 H}{dT_1^2} = K_a H - K_v \frac{dH}{dT_1} - K_G^2 \overline{G}^2,$$

$$H(T_1 = 0) = 1, \quad \frac{dH}{dT_1}(T_1 = 0) = 0,$$

$$H(T_1 = 1) = 1 + K_h \Delta H_0, \quad \frac{dH}{dT_1}(T_1 = 1) \to 0;$$

at the 2nd stage

$$K_h \frac{d\Delta H}{dT} = -2(K_h \Delta H + H_g)\frac{dL}{dT} - \frac{dH_g}{dT},$$

$$(K_h \Delta H + H_g) L^2 \frac{d^2 L}{dT^2} = K_w \overline{\rho} \overline{w}^2 \Delta H K_h - \lambda_{hy}(K_h \Delta H + H_g) L \left(\frac{dL}{dT}\right)^2,$$

$$\Delta H(T = 0) = \Delta H_0, \quad L(T = 0) = L_{ep}, \quad H(T = 0) = 1 + K_h \Delta H_0,$$

$$\frac{dH}{dT}(T=0) = 0, \qquad \frac{dL}{dT}(T=0) = K_L, \qquad \frac{dL}{dT}(L=1) \to 0,$$

where

$$K_G = \frac{M_G \Delta t_c}{\rho S_{ep} h_{ep}}, \qquad K_a = \frac{\Delta t_c^2}{h_{ep}}(a_{ep} - g), \qquad K_v = \frac{\nu \Delta t_c \Pi_{ep}}{h_{ep} S_{ep}},$$

$$K_h = \frac{h_{pr}}{h_{ep}}, \qquad K_w = M_w^2 \frac{h_{pr} M_t^2}{h_{ep} L_0^2}, \qquad K_L = \frac{G_c(t = \Delta t_c) M_t}{\rho S_{ep} L_0}.$$

Based on condition $K_G \equiv K_w \equiv K_L \equiv 1$ we obtain the scale:

$$M_t = \frac{\rho S_{ep} L_0}{G_c(t = \Delta t_c)} = \frac{L_0}{R'(0)},$$

$$M_G = \frac{\rho S_{ep} h_{ep}}{\Delta t_c},$$

$$M_w = \frac{h_{ep} G_c^2(t = \Delta t_c)}{h_{pr} \rho^2 S_{ep}^2} = \frac{h_{ep}}{h_{ep}} R'(0).$$

Therefore the final criterial form of the calculation model has the form:
at the 1st stage

$$\frac{dH}{dT_1} = -\overline{G}, \tag{19}$$

$$H \frac{d^2 H}{dT_1^2} = K_a H - K_v \frac{dH}{dT_1} - \overline{G}^2, \tag{20}$$

$$H(T_1 = 0) = 1, \qquad \frac{dH}{dT_1}(T_1 = 0) = 0, \tag{21}$$

$$H(T_1 = 1) = 1 + K_h \Delta H_0, \qquad \frac{dH}{dT_1}(T_1 = 1) \to 0; \tag{22}$$

at the 2nd stage

$$K_h \frac{d\Delta H}{dT} = -2(K_h \Delta H + H_g)\frac{dL}{dT} - \frac{dH_g}{dT}, \tag{23}$$

$$(K_h \Delta H + H_g) L^2 \frac{d^2 L}{dT^2} = \overline{\rho w}^2 \Delta H K_h - \lambda_{hy}(K_h \Delta H + H_g) L \left(\frac{dL}{dT}\right)^2, \tag{24}$$

$$\Delta H(T=0) = \Delta H_0, \qquad L(T=0) = L_{ep}, \qquad H_g(T=0) = 1 + K_h \Delta H_0, \tag{25}$$

$$\frac{d\Delta H}{dT}(T=0) = \frac{dH_g}{dT}(T=0) = 0, \qquad \frac{dL}{dT}(T=0) = 1, \qquad \frac{dL}{dT}(L=1) \to 0, \qquad (26)$$

$$H_g(t) = f_p(R), \qquad H_r(L=1) = h_к/h_{ер}. \qquad (27)$$

Condition of the industrial site flooding

$$\Delta H > 1. \qquad (28)$$

## Analysis of the model calculation

1. The calculation mathematical model represents a system of non-linear differential equations without trivial analytical solutions. Therefore the integration was made by a well-known Runge-Kutta method with relative calculation error less than 1%.

2. The main determined initial data of the calculation modeling were [1-3]: the maximal responce of the earthquake acceleration at the epicenter of 2.0g at depth of 24 km; conservative duration of shocks at the epicenter of 200s, and the distance from the epicenter to the Fukushima-Daiichi NPP industrial site was 160km (by some sources the shocks duration was no greater than 100s, and distance to the epicenter was 180 km); the height of the NPP industrial site over the sealevel was 10 m.

The undetermined initial data was: the area of the earthquake epicenter, the bottom relief from the epicenter to the industrial site; the wind with the same direction speed; hydrodynamic resistance coefficient. During the calculations the bottom relief was approximated and relative size of the earthquake epicenter was defined as 0.05. In order to determine of the deduced hydrodynamic resistance coefficient $\lambda_{hy}$ the variation calculations were made which deduced from the condition $\Delta H=1.5$ the value 1.8 (fig.2). The influence of the wind of the same direction was neglected.

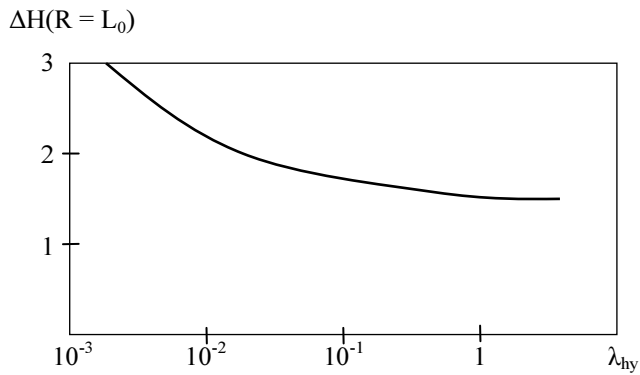

Fig.2. Dependency of a relative tsunami height near the shore on deduced hydrodynamic resistance coefficient.

3. The main results of the model calculation with average by travel section propagation speed and relative tsunami height are given at fig.3. At the initial stage with the earthquake excitation the values of speed and height of the tsunami abruptly increase. Further on under influence of dissipative processes of hydrodynamic resistance the speed of the wave propagation steadily reduces. As the wave approaches the water break there is an abrupt deceleration. The change in the wave speed and the depth reduction on the propagation path also determine the profile of the relative tsunami height change which reaches 15 m in front of the wave break.

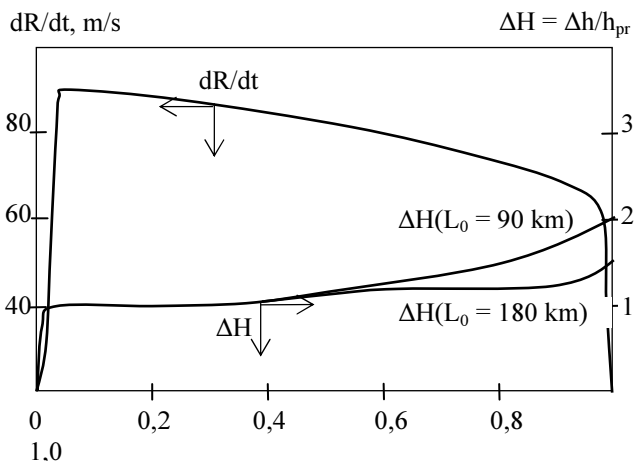

Fig.3. The main results of the calculation modeling of the average speed and relative tsunami height caused by 2.0 g earthquake.

4. Based on the criterial form of the tsunami hydrodynamical model analysis and variation calculations a significant dependency of the tsunami height near the shore on the following criterion was found:

$$C_1 = \frac{a_{ep}\Delta t_c h_{ep}}{\nu \lambda_{hy}}; \qquad C_2 = \frac{L_{ep}}{L_0}; \qquad C_3 = \frac{h_{\kappa}}{h_{ep}}. \qquad (29)$$

In particular with distance from the epicenter to the shore two times less than to the Fukushima-Daiichi NPP industrial site (approximately the location of the Onagawa NPP), the height of the wave may reach 20m (see fig.3). These results of the model calculations and also variation calculation on determination of $\lambda_{hy}$ (see fig.2) confirm significant influence of dissipative hydrodynamic losses on the basic characteristics of the tsunami.

The project and after-project (TEPCO-JSCE) estimates of the maximal possible tsunami height at the Fukushima-Daiichi NPP did not fully consider influence of the criterion (29), that finally led to underestimate of these estimates. On the other hand the proposed hydrodynamic model of the NPP industrial site flooding by a tsunami does not require implementation of not sufficiently justified 300%-500% redundancy in a tsunami wave height estimate, proposed by IAEA experts.

5. The main limitations of the proposed hydrodynamic model of the industrial site flooding are connected to the assumptions made on estimation of separate initial data (the earthquake epicenter size, the bottom relief, dissipative losses and other parameters) and also to averaging of the wave propagation speed by the square of the propagation section. These problems solution is a perspective for the proposed model improvement.

consequences of Fukushima NPP analysis for method method for heavy nuclear emergencies prevention – Chernobyl 2012, IPS NPP NASU, 280 pages.)

2. Билей Д.В., Ващенко В.Н., Злочевский В.В., Погосов А.Ю., Скалозубов В.И., Шавлаков А.В. Опыт АЭС Фукусима-1 для повышения экологической безопасности атомной энергетики Украины / монография / К: Государственная экологическая академия последипломного образования, 2012. – 194 с. (Biley D.V., Vashcenko V.N., Zlochevskiy V.V., Pogosov A.Yu, Skalozubov V.I., Shavlakov A.V. Application of Fukushima-1 NPP experience for Ukraine nuclear safety improvement / Kyiv: State Ecological Academy for post-graduate education, 2012, 194 pages)

3. Скалозубов В.И., Ващенко В.Н., Гоблая Т.В., Гудима А.А., Герасименко Т.В., Козлов И.Л. Повышение экологической безопасности атомной энергетики Украины в постфукусимский период / монография / K:2013. – 125с (Skalozubov V.I., Vaschenko V.N., Goblaia T.V., Gudyma A.A., Gerasimenko T.V., Kozlov I.L. Improvement in environmental safety of nuclear power in Ukraine in post-Fukushima period. Kyiv:2013, 125 pages)